\documentclass[aps,prc,twocolumn,groupedaddress,showpacs]{revtex4}
\usepackage{epsf}
\usepackage{dcolumn}

\def\bea {\begin{eqnarray}}
\def\eea {\end{eqnarray}}
\def\be {\begin{equation}}
\def\ee {\end{equation}}
\def\ben{\begin{enumerate}}
\def\een{\end{enumerate}}
\def\bi{\begin{itemize}}
\def\ei{\end{itemize}}
\def\ie{{\it i.e.}}
\def\viz{{\it viz.}\ }

\def\etal{{\it et al.}}
\def\F{{\cal F}}
\def\prl {Phys. Rev. Lett.\ }
\def\pl {Phys. Lett.\ }
\def\pr {Phys. Rev.\ }
\def\np {Nucl. Phys.\ }
\def\GV{G_{\mbox{\tiny V}}}

\def\DRV{\Delta_{\mbox{\tiny R}}^{\mbox{\tiny V}}}
\def\mA{m_{\mbox{\tiny A}}}
\def\mZ{m_{\mbox{\tiny Z}}}
\def\hyphen{{\mbox{-}}}
\newcommand{\sfrac}[2]{\mbox{\small{$\frac{#1}{#2}$}}}
\def\2p{|2p\rangle }
\def\4p2h{|4p\hyphen 2h\rangle }
\def\6p4h{|6p\hyphen 4h\rangle }

\begin{document} 
\title{Calculated Corrections to Superallowed
Fermi Beta Decay: New Evaluation of the Nuclear-Structure-Dependent Terms }
\author{I.S. Towner\footnote{Present address: Department of Physics,
Queen's University, Kingston, Ontario K7L 3N6, Canada} and J.C. Hardy}
\affiliation{Cyclotron Institute, Texas A \& M University,                    
College Station, Texas  77843}
\date{\today} 
\begin{abstract} 

The measured $ft$-values for superallowed $0^{+} \rightarrow 0^{+}$ nuclear
$\beta$-decay can be used to obtain the value of the vector coupling constant
and thus to test the unitarity of the Cabibbo-Kobayashi-Maskawa matrix.  An
essential requirement for this test is accurate calculations for the radiative
and isospin symmetry-breaking corrections that must be applied to the experimental
data.  We present a new and consistent set of calculations for the
nuclear-structure-dependent components of these corrections.  These new results do
not alter the current status of the unitarity test -- it still fails by more than two
standard deviations -- but they provide calculated corrections for eleven new
superallowed transitions that are likely to become accessible to precise measurements
in the future.  The reliability of all calculated corrections is explored and an
experimental method indicated by which the structure-dependent corrections can be
tested and, if necessary, improved.  

\end{abstract} 


\maketitle


\section{Introduction}
\label{s:intro}

Superallowed $0^{+} \rightarrow 0^{+}$ nuclear $\beta$-decay depends
uniquely on the vector part of the weak interaction.  When it occurs
between $T=1$ analog states, a precise measurement of the transition
$ft$-value can be used to determine $\GV$, the vector coupling constant.
This result, in turn, yields $V_{ud}$, the up-down element of the
Cabibbo-Kobayashi-Maskawa (CKM) matrix.  At this time, it is the key ingredient 
in one of the most exacting tests available of the unitarity of the CKM matrix, 
a fundamental pillar of the minimal Standard Model.  

Currently, there is a substantial body of precise $ft$-values determined for
such transitions and the experimental results are robust, most input data having
been obtained from several independent and consistent measurements \cite{TH98,Ha90}.
In all, $ft$-values have been determined for nine $0^{+} \rightarrow 0^{+}$
transitions to a precision of $\sim 0.1\%$ or better.  The decay parents -- $^{10}$C,
$^{14}$O,$^{26m}$Al,$^{34}$Cl,$^{38m}$K,$^{42}$Sc,$^{46}$V,$^{50}$Mn and
$^{54}$Co -- span a wide range of nuclear masses; nevertheless, as anticipated
by the Conserved Vector Current hypothesis, CVC, all nine yield consistent
values for $\GV$, from which a value 
of
\be
V_{ud} = 0.9740 \pm 0.0005
\label{Vud}
\ee
is derived.  The unitarity test of the CKM
matrix, made possible by this precise value of $V_{ud}$, fails by more than two
standard deviations \cite{TH98}: \viz
\be
V_{ud}^2 + V_{us}^2 + V_{ub}^2 = 0.9968 \pm 0.0014.
\label{unitarity}
\ee
In obtaining this result, we have used the Particle Data Group's \cite{PDG00}
recommended values for the much smaller matrix elements, $V_{us}$ and $V_{ub}$.
Although this deviation from unitarity is not completely definitive statistically, it is 
also supported by recent, less precise results from neutron decay \cite{Ab02}.  If the
precision of this test can be improved and it continues to indicate non-unitarity,
then the consequences for the Standard Model would be far-reaching.  

The potential impact of definitive non-unitarity has led to considerable recent activity, both
experimental and theoretical, in the study of superallowed $0^{+} \rightarrow 0^{+}$
transitions, with special attention being focussed on the small correction terms that must
be applied to the experimental $ft$-values in order to extract $\GV$.  Specifically,
$\GV$ is obtained from each $ft$-value via the 
relationship \cite{TH98}
\begin{equation}
\F t \equiv ft (1 + \delta_R )(1 - \delta_C ) =  
\frac{K}{2 \GV^2 (1 + \DRV )} ,
\label{Ft}
\end{equation}
\noindent with
\begin{eqnarray}
K/(\hbar c)^6 & = & 2 \pi^3 \hbar \ln 2 / (m_e c^2)^5
\nonumber \\
& = & ( 8120.271 \pm 0.012) \times 10^{-10} {\rm GeV}^{-4}
{\rm s},
\label{K}
\end{eqnarray}
\noindent where $f$ is the statistical 
rate function, $t$ is the partial half-life
for the transition, $\delta_C$ is the isospin-symmetry-breaking 
correction, $\delta_R$ is
the transition-dependent part of the radiative 
correction and $\DRV $ is the transition-independent
part.  Here we have also defined $\F t$ as the ``corrected"
$ft$-value.

It is now convenient to separate the radiative 
correction into two terms:
\be
\delta_R = \delta_R^{\prime} + \delta_{NS}
\label{deltaR}
\ee
where the first term, $\delta_R^{\prime}$, is a 
function of the electron's energy
and the charge of the daughter nucleus, $Z$; it therefore 
depends on the particular nuclear
decay, but is {\it independent} of nuclear structure. 
The second term, $\delta_{NS}$, is discussed more fully in
Sec.~\ref{ss:newdns} but its evaluation,
like $\delta_C$,  
depends on the details of nuclear
structure.  To emphasize the different sensitivities of the 
correction terms we re-write
the expression for $\F t$ as
\be
\F t \equiv ft (1 + \delta_R^{\prime} )(1 + \delta_{NS} - \delta_C ) 
\label{Ftnew}
\ee
where the first correction in brackets is 
independent of nuclear structure, while
the second incorporates the structure-dependent terms.  The term
$\delta_R^{\prime}$ has been calculated from 
standard QED, and is currently
evaluated to order $Z\alpha^2$ and estimated 
in order $Z^2\alpha^3$ \cite{Si87,JR87}; its
values are around 1.4\% and can be considered very reliable.  The
structure-dependent terms, $\delta_{NS}$ and $\delta_C$, have also
been calculated in the past but at various times over three decades
and with a variety of different nuclear models.  Their uncertainties
are larger.  This paper specifically addresses these correction terms
with a view to reducing their uncertainties.
 
Though depending on the nuclear shell-model, calculations for $\delta_{NS}$
and $\delta_C$ have been carefully linked to
other related observables such as the neutron and proton 
binding energies, the $b$- and $c$-coefficients
in the Isobaric Multiplet Mass Equation (IMME), and the 
non-analog $0^{+} \rightarrow 0^{+}$
transition rates (see, for example, \cite{OB85,Ha94,THH77}).   
Given this linking to observables and the more general success
of the shell model in this mass region, 
calculations of $\delta_{NS}$ and $\delta_C$ should 
also be rather reliable.  Nevertheless, conservative uncertainties
have been applied -- they are of 
order 0.1\% ({\it i.e.} $\sim 10\%$ of their own
value) -- and these become major contributors to the overall 
uncertainty on the unitarity test.  To illustrate: 
the uncertainty obtained for $V_{ud}$ in Eq.(\ref{Vud})is $\pm 0.0005$;  
the contributions to this uncertainty are 0.0001 from
experiment, 0.0001 from $\delta_R^{\prime}$, 
0.0003 from $\delta_C  -  \delta_{NS}$, and
0.0004 from $\DRV$.  If the unitarity test is to be 
sharpened, then the most pressing objective must be to
reduce the uncertainties on $\DRV$ and ($\delta_C  -  \delta_{NS}$).
The latter is clearly the  
most important area where nuclear physics can play an critical role.
There is considerable activity, both experimental
and theoretical, now underway in probing these nuclear-structure-dependent 
corrections with a view to
reducing the uncertainty that they introduce into the 
unitarity test.  

Since the goal of experiments will generally be to test and constrain the calculated
structure-dependent corrections, an important first step is to have a set of
consistent calculations that apply both to the nine well-known
transitions already used for the unitarity test and to possible new cases yet
to be studied.  In what follows, we present new calculations of
$\delta_C$ and $\delta_{NS}$, in which consistent model spaces and
approximations have been used for both correction terms and for a
large repertoire of superallowed transitions, new and old.  These
will provide a consistent standard for future experimental comparison.

\section{Theoretical corrections to superallowed decays}
\label{s:theo}

As described in the introduction, there are four theoretical
correction terms involved in extracting $V_{ud}$ from experimental
$ft$-values: the radiative corrections that are independent of
nuclear structure ($\delta_R^{\prime}$ and $\DRV$), the
nuclear-structure-dependent radiative correction ($\delta_{NS}$)
and the isospin-symmetry-breaking correction ($\delta_C$).  Though
we will later present new calculations of the last two, in this
section we present an overview of all four terms.  This overview
is placed in the context of a unitarity test that has failed
by more than two standard deviations.  In particular, we assess whether
the failure to meet unitarity can be removed by plausible adjustments
in these calculated corrections.  What changes would it take to
restore unitarity?  For example, $\delta_R^{\prime}$, would have to be
shifted downwards by 0.3\% (\ie \  as much as one-quarter of its
current value) for {\it all nine} currently well-measured nuclear
transitions; or ($\delta_{C} - \delta_{NS}$) would have to be shifted
upwards by 0.3\% (over one-half their value), for {\it all nine} cases;
or some combination of the two.  We will argue that such shifts are
very improbable. 

\subsection{Radiative corrections independent of nuclear structure}
\label{ss:rc}

The radiative correction comprises a 
transition-dependent term, $\delta_R$, and a transition-independent
term, $\DRV$.  The transition-dependent term is further divided
into $\delta_R^{\prime}$, which does not
depend on nuclear structure, and $\delta_{NS}$, which is
structure dependent.  We consider first the 
structure-independent terms, which are written:
\bea
\delta_R^{\prime} & = & \frac{\alpha}{2 \pi} \left [
\overline{g}(E_m)
+ \delta_2 + \delta_3  \right ] ,
\label{drDR}\\
\DRV & = & \frac{\alpha}{2 \pi} \left [ 4 \ln
(\mZ
/m_p )
+ \ln (m_p / \mA )
+ 2 C_{\rm Born} \right ] + \cdots ,
\nonumber
\eea
where the ellipses represent further
small
terms
of order
0.1\%.  In these equations, $E_m$ is the
maximum
electron energy
in beta decay, $\mZ$ the $Z$-boson mass, $m_p$ the proton mass,
$\mA$ the
$a_1$-meson mass, and
$\delta_2$ and $\delta_3$ the order-$Z
\alpha^2$
and
-$Z^2 \alpha^3$
contributions respectively.  The function
$g(E_e,E_m)$, which depends on the electron
energy, was first defined by Sirlin \cite{Si67} as
part of the order-$\alpha$ universal photonic
contribution arising from the weak vector current;
it is here
averaged
over
the
electron spectrum to give $\overline{g}(E_m)$. 
Finally, the term $C_{\rm Born}$ 
comes from the
order-$\alpha$ axial-vector photonic
contributions. 

Calculated values for all three components of
$\delta_R^{\prime}$ are given in Table~\ref{radct}.  There
have been two independent calculations 
\cite{Si87,JR87} of both $\delta_2$ and
$\delta_3$; they are completely consistent with
one another if proper account is taken of
finite-size effects in the nuclear charge
distribution.  The
values listed in Table~\ref{radct} are our
recalculations \cite{Ha90} using the formulas
of Sirlin \cite{Si87} but incorporating a Fermi
charge-density distribution for the nucleus.  Note
that we have followed Sirlin in assigning an
uncertainty equal to $(\alpha /2\pi ) \delta_3$ as an
estimate of the error made in stopping the
calculation at that order. 

\begin{table}
\begin{center}
\caption{Calculated nucleus-dependent radiative
correction,
$\delta_R$, in percent units, and the component
contributions
as identified in Eq.\ (\protect\ref{drDR}).
\label{radct} }
\vskip 1mm
\begin{ruledtabular}
\begin{tabular}{lcccc}
& \multicolumn{1}{c}{$\frac{\alpha}{2\pi } \overline{g} (E_m)$}
& \multicolumn{1}{c}{$\frac{\alpha}{2\pi } \delta_2$}
& \multicolumn{1}{c}{$\frac{\alpha}{2\pi } \delta_3$}
& \multicolumn{1}{c}{$\delta_R^{\prime}$} \\[1mm]
\hline 
& & & &  \\[-3mm]
$T_z = -1$: & & & & \\
~~ $^{10}$C  & 1.468    & 0.182       &  0.005      & 1.65(1)   \\ 
~~ $^{14}$O  & 1.286    & 0.227       &  0.008      & 1.52(1)   \\   
~~ $^{18}$Ne & 1.204    & 0.268       &  0.013      & 1.48(1)   \\ 
~~ $^{22}$Mg & 1.121    & 0.305       &  0.018      & 1.44(2)   \\ 
~~ $^{26}$Si & 1.055    & 0.338       &  0.024      & 1.42(2)   \\ 
~~ $^{30}$S  & 1.005    & 0.363       &  0.030      & 1.40(3)   \\ 
~~ $^{34}$Ar & 0.963    & 0.392       &  0.037      & 1.39(4)   \\ 
~~ $^{38}$Ca & 0.928    & 0.417       &  0.044      & 1.39(4)   \\ 
~~ $^{42}$Ti & 0.906    & 0.449       &  0.053      & 1.41(5)   \\[5mm] 
$T_z = 0 $: & & & & \\
~~ $^{26m}$Al & 1.110   & 0.325       &  0.021      & 1.46(2)   \\ 
~~ $^{34}$Cl & 1.002    & 0.388       &  0.034      & 1.42(3)   \\ 
~~ $^{38m}$K & 0.964    & 0.413       &  0.041      & 1.42(4)   \\ 
~~ $^{42}$Sc & 0.939    & 0.448       &  0.049      & 1.44(5)   \\ 
~~ $^{46}$V  & 0.903    & 0.468       &  0.057      & 1.43(6)   \\ 
~~ $^{50}$Mn & 0.873    & 0.494       &  0.065      & 1.43(7)   \\ 
~~ $^{54}$Co & 0.843    & 0.507       &  0.073      & 1.42(7)   \\ 
~~ $^{62}$Ga & 0.805    & 0.567       &  0.091      & 1.46(9)   \\
~~ $^{66}$As & 0.791    & 0.589       &  0.100      & 1.48(10)  \\
~~ $^{70}$Br & 0.777    & 0.609       &  0.110      & 1.50(11)  \\
~~ $^{74}$Rb & 0.763    & 0.627       &  0.120      & 1.51(12)  \\
\end{tabular}
\end{ruledtabular}
\end{center}
\end{table}

To assess the changes in $\delta_R^{\prime}$ that would be
required in order to restore unitarity, it is helpful
to rewrite Eq.\ 
(\ref{drDR}) in terms of the typical values taken
by its components:  \viz 
\be
\delta_R^{\prime} \simeq 1.00 + 0.40 + 0.05 \% ,
\label{tdr}
\ee
If the failure to obtain
unitarity
in the
 CKM matrix with
$V_{ud}$ from nuclear beta decay is due to the
value
of
this term alone,
then $\delta_R^{\prime}$ must be reduced to 1.1\%.  This
is not
likely.
The leading term, 1.00\%, involves standard
QED and
is
well
verified.  The order-$Z\alpha^2$ term, 0.40\%,
while
less
secure
has been calculated twice \cite{Si87,JR87}
independently, with
results in accord.

Taking a similar approach for the
nucleus-independent radiative
correction, we write

\be
\DRV = 2.12 - 0.03 + 0.20 + 0.1\%
~~\simeq~~
2.4\% ,
\label{tdrv}
\ee

\noindent of which the first term, the
leading
logarithm,
is
unambiguous.
Again, to achieve unitarity of the
CKM matrix,
$\DRV$
would have to be 
reduced to 2.1\%: \ie\, all terms other
than the
leading
logarithm must
sum to zero.  This also seems unlikely.
The adopted value of the nucleus-independent
radiative
correction has
been set at \cite{Si94}

\be
\DRV = (2.40 \pm 0.08) \% .
\label{DRV}
\ee

\noindent Note this value differs slightly (but
within
errors) from an 
earlier value \cite{MS86} because of the decision
by Sirlin
\cite{Si94} to centre the cut-off parameter
$\mA$, where $(m_{a_1}/2) \leq \mA \leq 2
m_{a_1}$, exactly at the $a_1$-meson mass
when
evaluating the axial contribution to the
radiative-correction
loop graph.  This range of possible values for $\mA$ is the dominant
contributor to the error in Eq.~(\ref{DRV}).

\subsection{The $\delta_{NS}$ correction}
\label{ss:newdns}

The nuclear-structure-dependent part of the radiative correction is
denoted $\delta_{NS}$.
Although for the superallowed transition we are discussing a purely
vector interaction between spin $0^{+}$ states, the
axial-vector interaction does play a role in the radiative corrections.
An axial-vector interaction may flip a nucleon spin and then be followed
by an electromagnetic interaction that flips it back again.
This axial contribution, denoted $C$, can be further divided into
two terms depending whether the weak and electromagnetic interactions
occur on the same nucleon or on two separate nucleons:
\bea
C & = & C_{{\rm Born}} + C_{NS} ,
\nonumber \\
\delta_{NS} & = & \frac{\alpha}{\pi} C_{NS} .
\label{Cdef}
\eea
Here $C_{{\rm Born}}$ refers to the Born graph in which the
axial-vector and electromagnetic interactions occur on the same nucleon.
This term is universal -- \ie \  the same in all nuclei -- so it is not
included in $\delta_{NS}$ but is placed in the nucleus-independent
radiative correction $\DRV$, see Eq.~(\ref{drDR}).  
The term, $C_{NS}$, refers to the case when
the axial-vector and electromagnetic interactions occur on different
nucleons.  The calculation of this term depends on the details
of nuclear structure.

In the earliest calculations of $\delta_{NS}$ \cite{JR90,BBJR92,To92}, 
the axial-vector and electromagnetic vertices 
were evaluated with free-nucleon coupling constants.  Yet there is
ample evidence in nuclear physics that coupling constants for
spin-flip processes are quenched in the nuclear medium.  Subsequently,
Towner \cite{To94} revised his earlier results \cite{To92} using
quenching factors that had been obtained previously \cite{To87,ASBH87,BW83}
from studies of weak and electromagnetic transitions in nuclei throughout
the region $10 \leq A \leq 54$.  These quenching factors depend weakly on
both mass and shell-model orbital.

\begin{table*}
\begin{center}
\caption{Shell-model calculations of the                       
nuclear-structure dependent component of the
radiative correction, $\delta_{NS}$.
The four components that are summed
to give $C_{NS}$ characterize the four electromagnetic
couplings:
os = orbital isoscalar,
ss = spin isoscalar,
ov = orbital isovector, and
sv = spin isovector.
\label{t:dnsnew}}
\vskip 1mm
\begin{ruledtabular}
\begin{tabular}{lddddddddd}
& & & & & & & & & \\[-3mm]
Parent   
 & \multicolumn{1}{c}{Unquenched}
 & \multicolumn{5}{c}{Quenched $C_{NS}$}
  & \multicolumn{1}{c}{$(q-1)\times$}
 & \multicolumn{2}{c}{$\delta_{NS}(\%)$} \\[1mm]
\cline{3-7} 
\cline{9-10} 
& & & & & & & & & \\[-3mm]
nucleus 
 & \multicolumn{1}{c}{$C_{NS}$}
 & \multicolumn{1}{c}{~~~~os} 
 & \multicolumn{1}{c}{~~~~ss} 
 & \multicolumn{1}{c}{~~~~ov} 
 & \multicolumn{1}{c}{~~~~sv} 
 & \multicolumn{1}{c}{~~~~total}
 & \multicolumn{1}{c}{$C_{{\rm Born}}({\rm free})$} 
 & \multicolumn{1}{c}{~~~Quenched}    
 & \multicolumn{1}{c}{~~~~~~~Adopted} \\[1mm]
\hline
& & & & & & & & & \\[-3mm]
$T_z = -1$: & & & & & & & & & \\
~~ $^{10}$C & -1.669 &   0.002  &  -0.283  &  -0.002  &  -1.065  &
       -1.348  &   -0.188  &  -0.357  &  -0.360(35)    \\
~~ $^{14}$O & -1.360 &  -0.008  &  -0.341  &  0.082  &  -0.782  &
       -1.049  &   -0.221  &  -0.295  &  -0.250(50)    \\
~~ $^{18}$Ne & -1.531 &  -0.011  &  -0.249  &  -0.119  &  -0.812  &
       -1.191  &   -0.210  &  -0.325  &  -0.290(35)    \\
~~ $^{22}$Mg & -1.046 &  -0.009  &  -0.222  &  -0.067  &  -0.497  &
       -0.796  &   -0.226  &  -0.237  &  -0.240(20)    \\
~~ $^{26}$Si & -0.986 &  -0.007  &  -0.224  &  -0.086  &  -0.424  &
       -0.741  &   -0.242  &  -0.228  &  -0.230(20)    \\
~~ $^{30}$S & -0.800 &  0.002  &  -0.287  &  0.020  &  -0.300  &
       -0.566  &   -0.257  &  -0.191  &  -0.190(15)    \\
~~ $^{34}$Ar & -0.770 &  0.014  &  -0.322  &  0.061  &  -0.272 &
       -0.519  &   -0.273  &  -0.184  &  -0.185(15)    \\
~~ $^{38}$Ca & -0.693 &  0.041  &  -0.358  &   0.091  &  -0.214  &
       -0.440  &   -0.288  &  -0.169  &  -0.180(15)    \\
~~ $^{42}$Ti & -1.011 &  -0.016  &  -0.181  &  -0.225  &  -0.354  &
       -0.776  &   -0.256  &  -0.240  &  -0.240(20)    \\[5mm]
$T_z = 0$: & & & & & & & & & \\
~~ $^{26m}$Al &  0.352 &  -0.007  &  -0.224  &   0.086  &   0.424  &
        0.279  &   -0.242  &   0.009  &  0.009(20)   \\
~~ $^{34}$Cl & -0.135 &   0.015  &  -0.333  &  -0.064  &   0.280  &
       -0.101  &   -0.273  &  -0.087  &  -0.085(15)    \\
~~ $^{38m}$K & -0.276 &   0.042  &  -0.363  &  -0.093  &   0.216  &
       -0.198  &   -0.288  &  -0.113  &  -0.100(15)    \\
~~ $^{42}$Sc &  0.472 &  -0.016  &  -0.182  &   0.228  &   0.358  &
        0.389  &   -0.256  &   0.031  &   0.030(20)  \\
~~ $^{46}$V &  0.101 &  -0.004  &  -0.197  &   0.099  &   0.198  &
        0.096  &   -0.263  &  -0.039  &   -0.040(7)    \\
~~ $^{50}$Mn &  0.054 &  -0.009  &  -0.184  &   0.104  &   0.152  &
        0.063  &   -0.270  &  -0.048  &   -0.042(7)    \\
~~ $^{54}$Co &  0.161 &  -0.013  &  -0.180  &   0.133  &   0.203  &
        0.144  &   -0.277  &  -0.031  &   -0.029(7)    \\
~~ $^{62}$Ga &  0.172 &   0.005  &  -0.289  &  -0.058  &   0.445  &
        0.103  &   -0.289  &  -0.043  &   -0.040(20)   \\
~~ $^{66}$As &  0.124 &   0.006  &  -0.291  &  -0.070  &   0.421  &
        0.066  &   -0.295  &  -0.053  &   -0.050(20)   \\
~~ $^{70}$Br &  0.077 &   0.009  &  -0.295  &  -0.083  &   0.401  &
        0.032  &   -0.301  &  -0.063  &   -0.060(20)   \\
~~ $^{74}$Rb &  0.155 &   0.009  &  -0.261  &   0.006  &   0.353  &
        0.106  &   -0.306  &  -0.046  &   -0.065(20)   \\

\end{tabular}
\end{ruledtabular}
\end{center}
\end{table*}

There is a further consideration.  The presence of
quenching also breaks the universality of the Born term,
$C_{{\rm Born}}$.  Writing the evaluation of $C_{{\rm Born}}$
with free-nucleon coupling constants as
$C_{{\rm Born}}({\rm free})$, then
$C_{{\rm Born}}({\rm quenched})$ can be written:

\bea
C_{{\rm Born}}({\rm quenched}) & = &
 q C_{{\rm Born}}({\rm free}) 
\label{CBorn}  \\
& = &C_{{\rm Born}}({\rm free}) + (q - 1)
C_{{\rm Born}}({\rm free}) ,
\nonumber
\eea

\noindent
where $q$ is the factor by which the product of the weak and electromagnetic coupling constants
is reduced in the medium relative to its free-nucleon value.  The first 
term in Eq.~(\ref{CBorn})  remains universal, while the second term
is now part of the nuclear-structure
dependence of the radiative correction.  Thus $\delta_{NS}$
is written

\be
\delta_{NS} = \frac{\alpha}{\pi}
\left \{ C_{NS}({\rm quenched}) + (q - 1)
C_{{\rm Born}}({\rm free}) \right \} .
\label{dNS}
\ee

We have calculated the $\delta_{NS}$ correction for a wide range
of nuclei with $0^+$ ($T = 1$) ground or isomeric states that
decay by superallowed $\beta$-emission; we used the shell
model with effective interactions as described in Appendix A.
Results for both quenched and unquenched coupling
constants are given in Table~\ref{t:dnsnew}.  All but the last
column in that table give the results from one particular calculation
for each parent nuclide.  (In most cases, two or three independent
calculations were performed for a single parent, each with a different
shell-model Hamiltonian.)  The last column lists the values we
adopt for $\delta_{NS}$: these values result from our assessment of the
quenched results from {\it all} calculations made for each decay -- not
just the ones shown in the previous columns -- with uncertainties chosen
to encompass the spread in the results from those calculations.

Extra details are also given in columns 3-6 of the table for the quenched
calculation since this is the version that we ultimately use in evaluating
$V_{ud}$.  With two-body operators there are two types of contributions:
those in which both interacting nucleons are in the valence model space,
and those in which one nucleon is in the valence space and one is
in the closed-shells core.  In the latter case a sum is required
over all the core nucleons.  The isospin structure 
of the operator is interesting to note: the weak interaction
contribution is isovector, while the electromagnetic contribution
is isoscalar or isovector.  The combined operator therefore
is either isovector or isotensor.  (An isoscalar combination 
is just proportional to the unit operator in isospin space and does
not induce a Fermi transition.)  Both the valence nucleons and those in the core 
contribute to the result for isovector operators, only the valence
nucleons contribute to the isotensor operators.

In Table~\ref{t:dnsnew} we show contributions to $C_{NS}$ from
the various components of the electromagnetic interaction:
orbital isoscalar (os),
spin isoscalar (ss),
orbital isovector (ov) and 
spin isovector (sv).  Note that the spin contributions are larger
than the orbital contributions.  Further, and more interesting, 
the isoscalar and isovector contributions are in phase when
the decaying nucleus has $T_z = -1$ and out of phase when the
decaying nucleus has $T_z = 0$.  This indicates that much larger
corrections are obtained in the $T_z = -1$ series than in the
$T_z = 0$ series.  If one looks at mirror transitions, this effect alone
contributes between $0.1 \%$ to $0.3 \%$ to a mirror asymmetry
in the $ft$-values.  Since current experiments aim at $ 0.1 \%$ 
accuracy, this effect might just be at the edge of detectability.

\subsection{Isospin symmetry-breaking corrections}
\label{ss:c}

Turning, next, to the isospin-symmetry breaking correction,
$\delta_C$, it too can be separated into two components:

\be
\delta_C = \delta_{C1} + \delta_{C2} .
\label{dc12}
\ee

\noindent
The first term, $\delta_{C1}$, arises from Coulomb and charge-dependent 
nuclear interactions that induce configuration mixing among the
$0^{+}$ state wave functions in both the parent and daughter nuclei.
Being charge dependent, this mixing serves to
break isospin symmetry between the analog parent and daughter states
of the superallowed transition.  The second term, $\delta_{C2}$, is due
to small differences in the single-particle neutron and proton
radial wave functions, which cause the radial overlap integral
of the parent and daughter nucleus to be less than unity.
Strictly speaking, these two aspects of the calculation of
$\delta_C$ cannot be separated, but in all but one calculation to date
(including those reported here) this division has been made.  (The exception
is the $(0 + 2 + 4)\hbar \omega$ large-basis shell-model
calculation of Navr\'{a}til \etal \cite{NBO97} for the lightest
superallowed emitter, $^{10}$C.)  This division is akin to
the division made in setting up a shell-model calculation, where
the configuration space is divided into a small, tractible valence
space and a remaining excluded space.  Then $\delta_{C1}$ 
arises from the charge-dependent mixing within the valence space,
while $\delta_{C2}$ represents the consequence of mixing
between configurations in the valence space with those in the
excluded space; this consequence being manifested by a change
in the single-particle radial wave function of the valence nucleons.

\subsubsection{The $\delta_{C1}$ correction}
\label{ss:newdc1}

If, in a shell-model calculation, the effective interaction is isospin
invariant, then the wave functions for the parent and daughter analogue states
are identical, and the square of the Fermi matrix element between
them (for isospin $T = 1$ states) is exactly $|M_F|^2 = 2$.  In
addition, beta transitions to all other $0^+$ states in the daughter are strictly
forbidden.  However, the addition of charge-dependent terms
to the effective interaction causes the breaking of analogue symmetry.
Under these conditions, the Fermi matrix element departs slightly from
its isospin-invariant value.  We write

\be
|M_F|^2 = 2 ( 1 - \delta_{C1}) .
\label{MF0}
\ee

\noindent
Also, with charge-dependent terms in the effective interaction,
the Fermi matrix elements to other non-analogue $0^{+}$ states
in the daughter are no longer exactly zero.  For example, there
could be small (usually less than 0.1\%) branches to those excited
$0^{+}$ states that are energetically accessible to beta-decay.
For the first excited (non-analog) $0^{+}$ state, we can write

\be
|M_F^1|^2 = 2 \delta_{C1}^1 .
\label{MF1}
\ee

\noindent

In a model calculation in which there
are only two basis states, the depletion of Fermi strength
in the ground-state transition is entirely picked up by the transition
to the excited non-analogue $0^{+}$ state.  Thus,

\be
\delta_{C1} = \delta_{C1}^1 .
\label{twostate}
\ee

\noindent
Further, if only two-state mixing is considered, the magnitude of
$\delta_{C1}$ is inversely proportional to the square of the excitation
energy of the excited $0^+$ state: \ie

\be
\delta_{C1} \propto \frac{1}{(\Delta E )^2} .
\label{dE2}
\ee

\noindent
For our calculations, in which a large number of basis states play a role,
Eqs.~(\ref{twostate}) and (\ref{dE2}) are no longer exact.  Even so, they
remain approximately true and continue to be a useful guide.  

Calculations of $\delta_{C1}$ turn out to be very sensitive
to the details of the model calculation.  This would be a very unfortunate
property if we were not able to adopt certain strategies that act to reduce
the model dependence considerably.  Because of the variation of $\delta_{C1}$ with
$(\Delta E )^2$ (see Eq.~(\ref{dE2})), it is important that
the isospin-independent Hamiltonian produce a good quality
spectrum of $0^+$ states.  Since this is not always possible to achieve
in the shell model, especially for nuclei near to closed shells, our first
strategy is to compensate for this by scaling the
calculated $\delta_{C1}$ values by a factor
$(\Delta E )^2_{{\rm theo}} /
(\Delta E )^2_{{\rm expt}}$, the ratio of the square of the
excitation energy of the first excited $0^+$ state in the
model calculation to that known experimentally.  The second
strategy we adopt to reduce the model dependence was first
used by Ormand and Brown\cite{OB85,OB89}.  We constrain the
charge-dependent part of the effective interaction to reproduce
other charge-dependent properties of the $0^+$ states, namely the coefficients
of the isobaric multiplet mass equation (IMME)\cite{Br98}.

\begin{table*}
\begin{center}
\caption{Shell-model calculations of the                       
isospin symmetry-breaking correction, $\delta_{C1}$.
\label{t:dc1new}}
\vskip 1mm
\begin{ruledtabular}
\begin{tabular}{lddddddd}
& & & & & & & \\[-3mm]
Parent
& \multicolumn{2}{c}{Measured IMME coefficients \protect\cite{Br98}}
& \multicolumn{1}{c}{$E_x(0^{+})$}
& \multicolumn{1}{c}{$E_x(0^{+})$}
& \multicolumn{3}{c}{$\delta_{C1}(\%)$} \\[1mm]
\cline{2-3}
\cline{6-8}
& & & & & & & \\[-3mm] 
nucleus  & \multicolumn{1}{c}{~~$b$}
& \multicolumn{1}{c}{~~~$c$}
& \multicolumn{1}{c}{~~expt}
& \multicolumn{1}{c}{~SM}
& \multicolumn{1}{c}{~~~unscaled}
& \multicolumn{1}{c}{~~~scaled}
& \multicolumn{1}{c}{~~~~~~~Adopted} \\[1mm]
\hline
& & & & & & & \\[-3mm]
$T_z = -1$: & & & & & & & \\
~~ $^{10}$C     &   -1.546  &    0.362  &   6.18    &
          11.05  &   0.002   &    0.007  &  0.010(10)  \\
~~ $^{14}$O    &   -2.493  &    0.337  &   6.59    &
          6.64   &    0.049  &    0.050  &  0.050(20)  \\
~~ $^{18}$Ne      &   -3.045(1)  &  0.347(1) &    3.63   &
          3.80   &   0.212   &     0.232 &  0.230(30)  \\
~~ $^{22}$Mg      &  -3.814(1)  &  0.315(1) &   6.24    &
          6.34   &   0.010   &   0.010   &  0.010(10)  \\
~~ $^{26}$Si     &  -4.535(2)  &  0.302(2) &   3.59    &
          4.96   &   0.022   &    0.042  &  0.040(10)  \\
~~ $^{30}$S      &  -5.185(2)  &  0.275(2) &   3.79    &
          3.86   &   0.186     &  0.193    &  0.195(30)  \\
~~ $^{34}$Ar      &   -5.777(2)  &  0.286(2) &   3.92    &
          3.91   &   0.031   &   0.030   &  0.030(10)  \\
~~ $^{38}$Ca     &   -6.328(3)  &  0.284(3) &   3.38    &
          3.21   &   0.026   &   0.023   &  0.020(10)  \\
~~ $^{42}$Ti    &  -6.712(3)  &  0.287(3) &   1.84    &
          3.60   &  0.065    &  0.249    &  0.220(100)  \\[5mm]
$T_z = 0 $: & & & & & & \\
~~ $^{26m}$Al     &  -4.535(2)  &  0.302(2) &  3.59     &
          4.96   &  0.022    &  0.041    &  0.040(10)  \\
~~ $^{34}$Cl     &  -5.777(2)  &  0.286(2) &   3.92   &
          3.91   &  0.103    &  0.103    &   0.105(20)  \\
~~ $^{38m}$K     &  -6.328(3)&  0.284(3) &   3.38    &
          3.21   &   0.099    &  0.089    &  0.100(20)  \\
~~ $^{42}$Sc     &  -6.712(3)  &  0.287(3) &   1.84    &
          3.60   &  0.019    &  0.072    &  0.060(30)  \\
~~ $^{46}$V   & -7.327(10) & 0.276(11) &  2.61     &
          3.92   &  0.043    &  0.097    &  0.095(20)  \\
~~ $^{50}$Mn   & -7.892(30) & 0.259(30) &  3.69     &
          4.23   &  0.048     &  0.063     &  0.055(20)  \\
~~ $^{54}$Co   & -8.519(25) & 0.276(25) &  2.56     &
          2.26   & 0.058     & 0.045     &  0.040(15)    \\
~~ $^{62}$Ga    &  -9.463(70)  & 
 0.265(25)\footnotemark[1] &  2.33     &
          2.26   &  0.350    &  0.330    &  0.330(40)  \\
~~ $^{66}$As     &  -9.95(15)   & 
 0.262(25)\footnotemark[1] &  2.17\footnotemark[2] &
          1.81   &  0.356    &  0.247    &  0.250(40)  \\
~~ $^{70}$Br     &  -10.48(23)  & 
 0.260(25)\footnotemark[1] &  2.01     &
          1.72   &  0.479    &  0.352    &  0.350(40)  \\
~~ $^{74}$Rb     &  -10.82(25)  & 
 0.258(25)\footnotemark[1] &  0.508    &
          0.523  &  0.122     &  0.129     &  0.130(60)  \\
\end{tabular}
\end{ruledtabular}
\footnotetext[1]{Estimated values extrapolated from a fit to $c$ coefficients
in $0^+$ states in $A = 4n + 2$ nuclei, $10 \leq A \leq 58$.} 
\footnotetext[2]{Estimated value taken to be an average of the excitation energies of
$0^+$ states in $^{62}$Zn and $^{70}$Se.}
\end{center}
\end{table*}

There are three ways in which charge dependence enters our shell-model
calculation.  First, the single-particle energies of the proton
orbits are shifted relative to those of the neutrons.  The amount of
shift is determined from the spectrum of single-particle states
in the closed-shell-plus-proton versus the closed-shell-plus-neutron
nucleus, where the closed shell is taken to be the nucleus used
as a closed-shell core in that particular shell-model calculation.
These single-particle shifts 
are taken from experiment and are not adjusted.  Second, a
two-body Coulomb interaction is added among the valence protons.
The strength of this interaction is adjusted so that the $b$-coefficient
of the IMME is exactly reproduced.  Third, we add a charge-dependent 
nuclear interaction by increasing all the $T = 1$ 
proton-neutron matrix elements by about $2\%$ relative to the
neutron-neutron matrix elements.  The precise amount of this
increment was determined by requiring that the $c$-coefficient of
the IMME be exactly reproduced.

For each of the nuclei appearing in the previous tables, we list
in Table~\ref{t:dc1new} the values of the corresponding measured
IMME coefficients, $b$ and $c$, together with the known excitation
energy, $E_{x}(0^+)$, of the lowest excited $0^+$ state in their daughters.
As already explained, all our shell-model calculations were adjusted
to reproduce exactly the values of $b$ and $c$, and any discrepancy
between the calculated and experimental values of $E_{x}(0^+)$ was
compensated for by scaling the calculated results for $\delta_{C1}$.  
As we did in Table~\ref{t:dnsnew}, Columns 5-7 of in this table give
the results from one particular calculation for each parent nucleus.
These columns  list the calculated $0^+$ excitation energy and
$\delta_{C1}$ values, both unscaled and scaled for any $E_{x}(0^+)$
discrepancy.  Finally, the eighth column gives the $\delta_{C1}$
values we adopt.  These values result from our assessment of the
results of {\it all} calculations made for each decay -- not just
the ones shown in columns 5-7 -- with uncertainties chosen to encompass
the spread in the results from those calculations and to include the
uncertainty in the IMME $b$- and $c$-coefficients. 

For the nuclei with $A \geq 38$ there are excited (non-analogue) $0^+$
states in the daughter nuclei that are accessible to beta decay.  Some
of the Fermi transitions to these states have also been measured
\cite{Ha94,DR85}.  In Table~\ref{t:dc1exnew} we list one set of
calculated $\delta_{C1}^1$ values, both unscaled and scaled, along
with the value of $\delta_{C1}^1$ we adopt based on the same assessment
as that described for Table~\ref{t:dc1new}.  As before, the assigned
errors reflect both the spread among the different calculations and the
uncertainties in the IMME coefficients.  The measured branching ratios
were then converted to $\delta_{C1}^1$ values (see Eq.~(\ref{MF1})), which
appear in the last column of the table.  With the possible exception of the
results for $^{50}$Mn, the agreement between theory and experiment is
entirely satisfactory.

\begin{table}
\begin{center}
\caption{Shell-model calculations for the square of the Fermi matrix
element to the first excited $0^+$ state, $\delta_{C1}^1$.                        
\label{t:dc1exnew}}
\vskip 1mm
\begin{ruledtabular}
\begin{tabular}{lcccd}
& & & & \\[-3mm]
Parent  
& \multicolumn{4}{c}{$\delta_{C1}^1(\%)$} \\
\cline{2-5}
& & & & \\[-3mm]
nucleus & unscaled & scaled & adopted
& \multicolumn{1}{c}{expt} \\[1mm]
\hline
& & & & \\[-3mm]
$T_z = 0 $: & & & & \\
~~ $^{38m}$K &   0.068 &  0.062 &  0.090 (30)  &  < 0.28  \footnotemark[1]  \\
~~ $^{42}$Sc &  0.007 &  0.029 &  0.020 (20)  &  0.040(9) \footnotemark[2] \\
~~ $^{46}$V &  0.020 &  0.046 &  0.035 (15)  &  0.053(5)  \footnotemark[1] \\
~~ $^{50}$Mn & 0.038 &  0.049 &  0.045 (20)  &  < 0.016  \footnotemark[1] \\
~~ $^{54}$Co & 0.049 & 0.038  &  0.040 (20)  &  0.035(5) \footnotemark[1] \\
~~ $^{62}$Ga &  0.089 &  0.084  & 0.085 (20) & \\
~~ $^{66}$As &  0.027 &  0.019 & 0.020 (20)  &  \\
~~ $^{70}$Br &  0.095 &  0.070 & 0.070 (20)  &  \\
~~ $^{74}$Rb &  0.045 &  0.047 & 0.050 (30)  &  \\
\end{tabular}
\end{ruledtabular}
\footnotetext[1]{From Hagberg \protect\etal \  
\protect\cite{Ha94}  }
\footnotetext[2]{From Daehnick and Rosa \protect\cite{DR85}, averaged
with earlier results.  }  
\end{center}
\end{table}

\subsubsection{The $\delta_{C2}$ correction}
\label{ss:newdc2}

The second isospin symmetry-breaking correction, $\delta_{C2}$,
accounts for the difference in radial forms between the proton
in the parent $\beta$-decaying nucleus and the neutron in the
daughter nucleus.  These radial forms are integrated together
and, if there were no difference between them, the integral would
just be the normalization integral of value one.  The departure
of the square of this overlap integral from unity corresponds
to $\delta_{C2}$.  There is a strong constraint on any
calculation of $\delta_{C2}$: the asymptotic forms of the radial
functions must be matched to the separation energies, $S_p$ and
$S_n$, where $S_p$ is the proton separation energy in the
parent nucleus and $S_n$ is the neutron separation energy in the
daughter nucleus.  These separation energies are well known and
and may be found in any atomic mass table.  It is the size of
the difference between $S_p$ and $S_n$ and whether or not the
radial wave functions have nodes that principally determine the
magnitude of $\delta_{C2}$.

Our calculations of this correction follow closely the methods 
described in our earlier work \cite{THH77}.  We use a Saxon-Woods
potential defined for a nucleus of mass $A$ and charge $Z+1$ as:
\bea
V(r) & = & - V_0 f(r) - V_s g(r) {\bf l}. \mbox{\boldmath$\sigma$}
+ V_C(r)
\nonumber \\
& & ~~~~~    - V_g g(r) - V_h h(r),
\label{VSW}
\eea
where
\bea
f(r) & = & \left \{ 1 + \exp \left ( (r-R)/a \right ) \right \}^{-1} ,  
\nonumber  \\
g(r) & = & \left ( \frac{\hbar}{m_{\pi} c} \right )^2 
\frac{1}{a_s r} \exp \left ( \frac{r - R_s}{a_s} \right )
\nonumber \\
& & ~~~~~ \times \left \{ 1 + \exp \left ( \frac{r - R_s}{a_s} \right )
\right \}^{-2} ,
\nonumber  \\
h(r) & = & a^2 \left ( \frac{df}{dr} \right )^2 ,
\nonumber  \\
V_C(r) & = & Z e^2 / r , ~~~~ {\rm for}~~ r \geq R_c
\nonumber  \\
 & = & \frac{Z e^2}{2 R_c} \left ( 3 - \frac{r^2}{R_c^2} \right )
 , ~~~~ {\rm for}~~ r < R_c ,
\label{Pot}
\eea
with $R = r_0 (A - 1)^{1/3}$ and
$R_s = r_s (A - 1)^{1/3}$.   Note that $g(r)$ is rendered dimensionless
through the use of the pion Compton wavelength, $\left ( \hbar /
m_{\pi} c \right )^2 = 2$ fm$^2$.   The first three terms in
Eq.~(\ref{VSW}) are the central, spin-orbit and Coulomb terms
respectively.  The fourth and fifth terms are additional surface
terms whose role we discuss shortly.
The parameters of the spin-orbit force
were fixed at standard values, $V_s = 7$ MeV, $r_s = 1.1$ fm and
$a_s = 0.65$ fm, leaving four parameters to be determined: $R_c$,
the radius of the Coulomb potential, and $V_0$, $r_0$ and $a$
characterizing the strength, range and diffuseness of the Saxon-Woods
potential.

To determine the radius of the Coulomb potential, $R_c$, we first 
obtained the charge mean-square radius, $\langle r^2 \rangle_{{\rm ch}}^{1/2}$,
of the decaying nucleus.  We used results from electron scattering
experiments \cite{De87}, which actually provide the charge radius of a stable
isotope of each element rather than the beta-decaying isotopes of interest here.
However, by examining the data on isotope shifts of charge radii we could make
corrections for this effect to arrive at radius values applicable to the
decaying nuclides; we enlarged the assigned error accordingly.
Our selected values of $\langle r^2 \rangle_{{\rm ch}}^{1/2}$ 
and their assigned errors are listed in Table~\ref{t:dc2new}.  To obtain
an appropriate value for $R_c$, two further adjustments are required
to the experimental values of $\langle r^2 \rangle_{{\rm ch}}^{1/2}$:
first, the finite size of the proton must be incorporated and second,
because the shell model uses $A$ single-particle coordinates rather
than ($A-1$) relative coordinates, a centre-of-mass correction must be applied.
With a Gaussian form for the proton single-particle density
and harmonic oscillator wave functions for the shell model,
the shell-model rms radius,
$\langle r^2 \rangle_{{\rm sm}}^{1/2}$,
relates to the experimentally measured rms radius via

\be
\langle r^2 \rangle_{{\rm ch}} = 
\langle r^2 \rangle_{{\rm sm}} + \frac{3}{2} \left (
a_p^2 - b^2/A \right ) ,
\label{r2sm}
\ee

\noindent
where $a_p = 0.694$ fm is the length parameter in the proton density
and $b$ is the length parameter of the harmonic oscillator,
approximately $b^2 = A^{1/3}$ fm$^2$. The Coulomb potential in
Eq.~(\ref{Pot}) is that of a uniformly charged sphere.  We match
the charge radius of this distribution with $\langle r^2 \rangle_{{\rm sm}}^{1/2}$
to determine the radius, $R_c$,

\be
R_c^2 = \frac{5}{3}
\langle r^2 \rangle_{{\rm sm}} .
\label{Rc2}
\ee

\begin{table*}
\begin{center}
\caption{Calculations of $\delta_{C2}$ with Saxon-Woods radial
functions, without parentage expansions, $\delta_{C2}^{I}$,
and with parentage expansions, $\delta_{C2}^{II}$, $\delta_{C2}^{III}$
and $\delta_{C2}^{IV}$.
\label{t:dc2new}}
\vskip 1mm
\begin{ruledtabular}
\begin{tabular}{lccccccc}
& & & & & & & \\[-3mm]
Parent   
 & \multicolumn{2}{c}{Radius parameters (fm)}  
 & & & & &\multicolumn{1}{c}{Adopted value} \\[1mm]
\cline{2-3} 
\cline{8-8} 
& & & & & & & \\[-3mm]
nucleus  
 & \multicolumn{1}{c}{$\langle r^2 \rangle_{{\rm ch}}^{1/2}$} 
& \multicolumn{1}{c}{$r_0$} &   
\multicolumn{1}{c}{$\delta_{C2}^{I}(\%)$} &
\multicolumn{1}{c}{$\delta_{C2}^{II}(\%)$} &
\multicolumn{1}{c}{$\delta_{C2}^{III}(\%)$} &
\multicolumn{1}{c}{$\delta_{C2}^{IV}(\%)$} &   
\multicolumn{1}{c}{$\delta_{C2}(\%)$} \\[1mm]
\hline
& & & & & & & \\[-3mm]
$T_z = -1$: & & & & & &  & \\
~~$^{10}$C   & 2.47(6)   & 0.931(66) & 0.132(10) & 0.167(12)      
  & 0.169(11) & 0.167(12)   &  0.170(15)    \\
~~$^{14}$O   & 2.74(4) &  1.244(32) & 0.217(11) & 0.270(12)      
  & 0.267(13) & 0.267(13)   &  0.270(15)  \\
~~$^{18}$Ne    & 3.00(3)  & 1.361(20)  & 0.251(6) & 0.386(9)      
  & 0.387(8) & 0.381(10)    &  0.390(10)   \\
~~$^{22}$Mg  & 3.05(4)   & 1.281(26) & 0.207(8)  & 0.249(9)     
  & 0.261(10) & 0.250 (8)   & 0.255(10)   \\
~~$^{26}$Si   & 3.10(3)   & 1.206(18) & 0.223(7) & 0.332(10)      
  & 0.327(11) & 0.323(10)   & 0.330(10)    \\
~~$^{30}$S    & 3.24(2)   & 1.223(13) & 0.812(15) & 0.728(15)      
  & 0.730(17) & 0.750(16)   & 0.740(20)  \\
~~$^{34}$Ar   & 3.33(3)   & 1.253(17) & 0.351(15) & 0.650(21)     
  & 0.610(26) & 0.556(19)   &  0.610(40)   \\
~~$^{38}$Ca   & 3.48(2) & 1.269(10) & 0.402(11) & 0.727(17)      
  & 0.674(18) & 0.596(12)   & 0.710(50)  \\
~~$^{42}$Ti   & 3.60(5) & 1.316(22)   & 0.359(14) & 0.563(26)      
 & 0.572(29) & 0.578(33)    &  0.555(40)  \\[5mm]      
$T_z = 0 $: & & & & & &  \\
~~$^{26m}$Al  & 3.04(2)   & 1.194(12) & 0.156(3)  & 0.231(5)     
 & 0.227(5) & 0.225(4)      & 0.230(10)   \\
~~$^{34}$Cl   & 3.39(2)   & 1.303(11) & 0.312(8)  & 0.557(11)      
 & 0.536(15) & 0.479(11)    & 0.530(30)   \\
~~$^{38m}$K    & 3.41(4)  & 1.245(21) & 0.299(18) & 0.540(28)     
 & 0.495(30) & 0.445(20)    & 0.520(40)   \\
~~$^{42}$Sc     & 3.53(5) & 1.301(22) & 0.278(11) & 0.435(20)     
 & 0.438(26) & 0.446(28)    & 0.430(30)   \\     
~~$^{46}$V  & 3.60(7)  & 1.285(31)  & 0.273(17) & 0.344(21)     
 & 0.341(22) & 0.322(18)    & 0.330(25)  \\     
~~$^{50}$Mn  & 3.68(7) &  1.260(30)  & 0.315(20) & 0.439(27)     
 & 0.455(33) & 0.438(28)    & 0.450(30)  \\     
~~$^{54}$Co  & 3.83(7) & 1.275(29)  & 0.376(22) & 0.578(34)     
 & 0.577(39) & 0.563(35)    & 0.570(40)  \\     
~~$^{62}$Ga  & 3.94(10)   & 1.271(42) & 1.31(11)  & 1.10(11)     
 & 1.07(11) & 1.01(8)       & 1.05(15)   \\
~~$^{66}$As   & 4.02(10)   & 1.264(41) & 1.32(12)  & 1.25(12)      
 & 1.18(14) & 1.07(8)     & 1.15(15)    \\
~~$^{70}$Br & 4.10(10)   & 1.264(39) & 1.43(13)  & 1.11(13)     
 & 1.03(14) & 0.85(6)     & 1.00(20)    \\
~~$^{74}$Rb       & 4.18(10) & 1.276(37) & 0.68(9)   & 1.51(14)     
 & 1.38(18) & 1.20(12)    & 1.30(40)    \\     
\end{tabular}
\end{ruledtabular}
\end{center}
\end{table*}

Finally, it remains to determine the parameters of the central potential,
$ V_0 f(r)$.  The diffuseness is fixed at the same value as that of
the spin-orbit potential, $a = 0.65$ fm, for all $A$ values except
the lightest, $A = 10$ and 14, for which we used $a = 0.55$ fm.  The well depth,
$V_0$, was adjusted case-by-case so that the asymptotic form of the wave
function exactly matched that required for the known separation energy, $S_p$.
With the well depth so fixed, we computed the radial wave functions
for all proton states bound in that potential and constructed the
charge density of the nucleus from the square of these functions:

\be
\langle r^2 \rangle_{{\rm sm}} =
\frac{1}{Z} \sum_{nlj} (2 j + 1)
\langle r^2 \rangle_{n l j} ,
\label{r2SW}
\ee

\noindent
where $2 j + 1$ is the occupancy of protons in each orbital, $n l j$,
and the sum is over the occupied orbitals.  Here

\be
\langle r^2 \rangle_{n l j} =
\int_0^{\infty} | R_{n l j}(r) |^2 r^4 dr /
\int_0^{\infty} | R_{n l j}(r) |^2 r^2 dr ,
\label{r2nlj}
\ee

\noindent
with $R_{n l j}(r)$ being the radial wave function of the proton with
quantum numbers, $n l j$.  We then determined the radius parameter of
the Saxon-Woods potential, $r_0$, by requiring the
$\langle r^2 \rangle_{{\rm sm}}^{1/2}$ computed from Eq.~(\ref{r2SW})
to match that determined from experimental
electron scattering, Eq.~(\ref{r2sm}).  The value of $r_0$ is also given
in Table~\ref{t:dc2new} and its error reflects the assigned error on
$\langle r^2 \rangle_{{\rm ch}}^{1/2}$.

In the shell model, the $A$-particle wave functions,
$| J_i T_i \rangle$ and
$| J_f T_f \rangle$, can be expanded into products of $(A-1)$-particle
wave functions $|\pi \rangle $ and single-particle functions
$| j \rangle $.  In terms of this expansion, the Fermi matrix
element is

\begin{eqnarray*}
\lefteqn{M_F = } \\
& & \sqrt{\frac{3}{2}} \langle T_f M_{T_f} 1 1 | T_i M_{T_i} \rangle \\
& & \times \left \{ \sum_{j \pi } U(1 \sfrac{1}{2} T_i T_{\pi}; \sfrac{1}{2} T_f)
S^{1/2}( i \{ | \pi ; j ) S^{1/2}( f \{ | \pi ; j )\Omega_j^{\pi} \right \}
\end{eqnarray*}

\be
\Omega_j^{\pi}  =  
\int_0^{\infty} R_{\pi j}^p(r) R_{\pi j}^n r^2 dr .
\label{exps}
\ee

\noindent
The expansion coefficients
$S^{1/2}( i \{ | \pi ; j )$ and
$S^{1/2}( f \{ | \pi ; j )$ are generalised fractional parentage
coefficients and represent the spectroscopic overlap of the $A$- and
$(A-1)$-particle wave functions.  The sum in Eq.~(\ref{exps}) is
over all parent states, $|\pi \rangle$, and all single-particle
orbitals active in the shell-model calculation.  Note that the
radial integrals,
$\Omega_j^{\pi}$, 
are labelled with $\pi$.  These integrals are evaluated with 
eigenfunctions of the Saxon-Woods potential whose well depth
is continually adjusted to match the separation energy
to that particular parent state.  If we do not allow the
proton and neutron radial functions, $R^p(r)$ and $R^n(r)$,
to vary with the parent states but fix their asymptotic forms
for all $j$ to the separation energy of the ground state of the
parent nucleus, then the sums over $\pi$ can be done analytically
and the computed value of $\delta_{C2}$ becomes independent
of the shell-model effective interaction.  Results of this calculation
are given in Table~\ref{t:dc2new} and labelled $\delta_{C2}^{I}$.
Results without this simplifying assumption are also given and labelled
$\delta_{C2}^{II}$.  These latter results depend on the effective
interaction but not strongly.  One reason for this is that 
in implementing Eq.~(\ref{exps}), we use experimental excitation
energies in the $(A-1)$ nucleus for the lowest-energy state of each
spin and parity.  The shell model is used to provide spectroscopic
amplitudes and the excitation energies of states in the $(A-1)$
nucleus relative to the lowest state of that spin and parity.
The difference between $\delta_{C2}^{I}$ and $\delta_{C2}^{II}$ 
indicates the role of
the parentage expansions.

So far, the two surface terms in Eq.~(\ref{Pot}) have not been included,
$V_g = 0$, $V_h = 0$.
It can be argued that the central part of the potential,
which in principle should be determined from some Hartree-Fock
procedure, should not be continually adjusted.  Rather, any alteration
should be to the surface part of the potential.  Thus, in this
method, we fix $V_0$ separately for protons and neutrons to match
the ground-state parent separation energies, $S_p$ and $S_n$.  For the
excited parent states of excitation energy, $E_x$, 
we adjust the strength of the surface term,
$V_g$ (keeping $V_h = 0$) so that the asymptotic forms match
the separation energies $S_p + E_x$ and $S_n + E_x$.
These results are listed in Table~\ref{t:dc2new} as $\delta_{C2}^{III}$.  

The second surface term, $h(r)$, is even more strongly peaked in the surface
than $g(r)$.  Thus our fourth method is the same as the third, 
except that it is the
second surface term, $V_h$, that was adjusted, keeping $V_g = 0$.
These results are listed in Table~\ref{t:dc2new} as $\delta_{C2}^{IV}$.  

On average, the method III values of $\delta_{C2}$ are about 2\% lower
than the method II values; and method IV values about 7\% lower than
the method II values.  These are not big differences.  The errors on
each individual entry of $\delta_{C2}$ in the Table~\ref{t:dc2new}
reflects only the error in this quantity due to the uncertainty in
the rms charge radius $\langle r^2 \rangle^{1/2}$.  Once again, as we have
done in previous tables, the values tabulated for $\delta_{C2}^{I}$,
$\delta_{C2}^{II}$, $\delta_{C2}^{III}$ and $\delta_{C2}^{IV}$ give the
results from one particular calculation for each parent nucleus.  Our adopted
$\delta_{C2}$ values result from our assessment of {\it all} multiple-parentage
calculations made for each decay -- not just those shown in the preceeding
three columns.  The error on our adopted value reflects not only the
uncertainty in the rms charge radius, but also the spread of results obtained
with different shell-model effective interactions and the different
procedures, II, III, and IV.

\subsection{Collected structure-dependent corrections: their reliability}
\label{ss:cpw}

Our adopted values for the three nuclear-structure-dependent corrections,
$\delta_{NS}$, $\delta_{C1}$ and $\delta_{C2}$ are collected in Table~\ref{t:adopt}.
Since their impact on the $ft$ values is in the combination ($\delta_C - \delta_{NS}$),
(see Eq.~(\ref{Ftnew})), where $\delta_C = \delta_{C1} + \delta_{C2}$, we list our results
for this combination with the individual errors added in quadrature.  Note that
in the combination ($\delta_C - \delta_{NS}$) all three corrections are in
phase with the exception of the small $\delta_{NS}$ values in the cases of
$^{26m}$Al and $^{42}$Sc.  For the nine nuclei for which precision $ft$ values
have been measured, $^{10}$C and $^{14}$O of the $T_z = -1$ series,
and $^{26m}$Al to $^{54}$Co of the $T_z = 0$ series, the nuclear-structure
correction ranges from a low of 0.26\% for $^{26m}$Al to a high of 0.72\% for
$^{38m}$K.  Of particular interest is that larger values are found at the
upper end of the $s,d$-shell in the $T_z = -1$ series and at the upper end of
the $p,f$-shell in the $T_z = 0$ series.  This is mainly due to the radial 
overlap correction, $\delta_{C2}$, which yields larger numerical
values whenever a single-particle orbital with a radial node
contributes importantly in the parentage expansions, such as the $2s_{1/2}$
orbital in the upper $s,d$-shell and the $2p_{3/2}$, $2p_{1/2}$
orbitals in the upper $p,f$-shell.

\begin{table}
\begin{center}
\caption{Adopted values for the three nuclear-structure dependent
corrections for superallowed Fermi $\beta$ decay.
\label{t:adopt}}
\vskip 1mm
\begin{ruledtabular}
\begin{tabular}{lllll}
& & & & \\[-3mm]
Parent    
 & \multicolumn{1}{c}{$\delta_{NS}(\%)$}     
 & \multicolumn{1}{c}{$\delta_{C1}(\%)$}     
 & \multicolumn{1}{c}{$\delta_{C2}(\%)$}     
 & \multicolumn{1}{c}{$\delta_{C}      
 - \delta_{NS}(\%)$} \\[1mm]
\hline
& & & & \\[-3mm]
$T_z = -1$: & & & & \\
$^{10}$C &  $ -0.360(35) $ & $ ~~0.010(10) $ & $ ~~0.170(15) $ & 
      $ ~~0.540(39) $ \\
$^{14}$O &  $ -0.250(50) $ & $ ~~0.050(20) $ & $ ~~0.270(15) $ &
      $ ~~0.570(56) $ \\
$^{18}$Ne&  $ -0.290(35) $ & $ ~~0.230(30) $ & $ ~~0.390(10) $ &
      $ ~~0.910(47) $ \\
$^{22}$Mg&  $ -0.240(20) $ & $ ~~0.010(10) $ & $ ~~0.255(10) $ &
      $ ~~0.505(24) $ \\
$^{26}$Si&  $ -0.230(20) $ & $ ~~0.040(10) $ & $ ~~0.330(10) $ &
      $ ~~0.600(24) $ \\
$^{30}$S &  $ -0.190(15) $ & $ ~~0.195(30) $ & $ ~~0.740(20) $ &
      $ ~~1.125(39) $ \\
$^{34}$Ar&  $ -0.185(15) $ & $ ~~0.030(10) $ & $ ~~0.610(40) $ &
      $ ~~0.825(44) $ \\
$^{38}$Ca&  $ -0.180(15) $ & $ ~~0.020(10) $ & $ ~~0.710(50) $ &
      $ ~~0.910(53) $ \\
$^{42}$Ti&  $ -0.240(20) $ & $ ~~0.220(100) $ & $ ~~0.555(40) $ &
      $ ~~1.015(110) $ \\[5mm]
$T_z = 0$: & & & & \\
$^{26m}$Al&  $~~0.009(20) $ & $ ~~0.040(10) $ & $ ~~0.230(10) $ &
      $ ~~0.261(24) $ \\
$^{34}$Cl&  $ -0.085(15) $ & $ ~~0.105(20) $ & $ ~~0.530(30) $ &
      $ ~~0.720(39) $ \\
$^{38m}$K &  $ -0.100(15) $ & $ ~~0.100(20) $ & $ ~~0.520(40) $ &
      $ ~~0.720(47) $ \\
$^{42}$Sc&  $~~0.030(20) $ & $ ~~0.060(30) $ & $ ~~0.430(30) $ &
      $ ~~0.460(47) $ \\
$^{46}$V &  $ -0.040(7) $ & $ ~~0.095(20) $ & $ ~~0.330(25) $ &
      $ ~~0.465(33) $ \\
$^{50}$Mn&  $ -0.042(7) $ & $ ~~0.055(20) $ & $ ~~0.450(30) $ &
      $ ~~0.547(37) $ \\
$^{54}$Co&  $ -0.029(7) $ & $ ~~0.040(15) $ & $ ~~0.570(40) $ &
      $ ~~0.639(43) $ \\
$^{62}$Ga&  $ -0.040(20)$ & $ ~~0.330(40) $ & $ ~~1.05(15) $ &
      $ ~~1.42(16)  $ \\
$^{66}$As&  $ -0.050(20)$ & $ ~~0.250(40) $ & $ ~~1.15(15) $ &
      $ ~~1.45(16)  $ \\
$^{70}$Br&  $ -0.060(20)$ & $ ~~0.350(40) $ & $ ~~1.00(20) $ &
      $ ~~1.41(21) $ \\
$^{74}$Rb&  $ -0.065(20)$ & $ ~~0.130(60) $ & $ ~~1.30(40) $ &
      $ ~~1.50(41)  $ \\
\end{tabular}
\end{ruledtabular}
\end{center}
\end{table}

There have been a number of previous calculations of $\delta_C$ but
only one of $\delta_{NS}$.  The latter was performed by one of the 
present authors \cite{To94} using the same techniques described here
but applied only to the nine well-known superallowed transitions and
with similar -- though different in detail -- shell-model calculations
to ours; the results for those transitions are very similar to the
present results, well within the error bars in all cases.

The more numerous results from previous $\delta_C$ calculations appear in
Table~\ref{t:previous}, where they are compared with our present results.
Four groups of authors have published values for $\delta_C$, the first
in 1973.  In the table, we present the most recent results from each group for
each transition.  The values in the first column are those calculated
previously by us, reported first in refs. \cite{THH77,TH73}
and then refined in more recent publications \cite{To89,Ha94}.  These
were based on the same methods as those used here: shell-model
calculations to determine $\delta_{C1}$, and full-parentage
expansions in terms of Woods-Saxon radial wave functions to
obtain $\delta_{C2}$.  Ormand and Brown, whose values \cite{OB95} for
$\delta_C$ appear in column 2, also employed the shell model for
calculating $\delta_{C1}$, but they derived $\delta_{C2}$ from a
self-consistent Hartree-Fock calculation.  Both of these independent
calculations for $\delta_C$ -- those in columns one and two -- reproduce
the measured coefficients of the relevant isobaric multiplet mass
equation, the known proton and neutron separation energies, and the
measured $ft$-values of the weak non-analogue $0^{+} \rightarrow
0^{+}$ transitions \cite{Ha94} where they are known.  The agreement of
these calculations with our new results is rather good, especially for the
well known nine.  In the cases of the less well known $T_z = -1$ nuclei
between $^{18}$Ne and $^{42}$Ti, the differences are in general larger, but
this reflects improvements to $sd$-shell calculations realized since 1973,
when the only previous calculations\cite{TH73} were published.

The other two previous calculations shown in the table provide a valuable
check that these $\delta_C$ values do not suffer from severe systematic effects. 
Sagawa, van Giai and Suzuki\cite{SVS96}
have added RPA correlations to a Hartree-Fock
calculation that incorporates charge-symmetry
and charge-independence breaking forces in the
mean-field potential to take account of isospin
impurity in the core; the correlations, in
essence, introduce a coupling
to the isovector monopole giant resonance.  The
calculation is not constrained, however, to
reproduce known separation energies.  In addition, the authors
themselves\cite{SVS96} admit that their HF+RPA calculations cannot
properly take account of pairing in open-shell nuclei; as a consequence,
the discrepancies between their values and the others for $^{34}$Cl and
$^{38m}$K is not considered significant.  Clearly the overall trend of the
shell-model-based calculations is well reproduced by these very-different
calculations, thus ruling out the possibility that the former had missed
significant core contributions. Finally, a large shell-model calculation
has been mounted for the $A=10$ case by 
Navr\'{a}til, Barrett and Ormand \cite{NBO97}.  This ``microscopic"
calculation of $\delta_C$ also supports the results of the more 
macroscopic calculations reported here and in columns 1 and 2.

\begin{table}
\begin{center}
\caption{Calculated values for the isospin symmetry-breaking
correction, $\delta_C$ in percent units.  Previous calculations
are compared with the present results. 
\label{t:previous}}
\vskip 1mm
\begin{ruledtabular}
\begin{tabular}{llllll}
& & & & & \\[-3mm]
Parent
& \multicolumn{1}{c}{Towner}
& \multicolumn{1}{c}{Ormand}
& \multicolumn{1}{c}{Sagawa}
& \multicolumn{1}{c}{Navr\'{a}til}
& \multicolumn{1}{c}{Present} \\
& & & & & \\[-3mm]
nucleus\
& \multicolumn{1}{c}{\& Hardy\footnotemark[1]}
& \multicolumn{1}{c}{\& Brown\footnotemark[2]}
& \multicolumn{1}{c}{{\it et al}\footnotemark[3]}
& \multicolumn{1}{c}{{\it et al}\footnotemark[4]}
& \multicolumn{1}{c}{work}  \\[1mm]
\hline
& & & & & \\[-3mm]
$T_z = -1$: & & & & & \\
~~$^{10}$C &  0.18(2)  & 0.15(9)  & 0.00  &  0.12  &  0.18(2) \\         
~~$^{14}$O &  0.28(3)\footnotemark[5]  & 0.15(9)  & 0.29  &        &  0.32(3)  \\
~~$^{18}$Ne &  0.45(3)  &  &  &  &  0.62(3)  \\
~~$^{22}$Mg &  0.35(3)  &  &  &  &  0.27(2)  \\
~~$^{26}$Si &  0.42(4)  &  &  &  &  0.37(2)  \\
~~$^{30}$S &  1.21(10)  &  &  &  &  0.94(4)  \\
~~$^{34}$Ar &  1.04(9)  &  &  &  &  0.64(4)  \\
~~$^{38}$Ca &  0.89(9)  &  &  &  &  0.73(5)  \\
~~$^{42}$Ti &  0.62(6)  &  &  &  &  0.78(11)  \\
$T_z = 0$: & & & & & \\
~~$^{26m}$Al& 0.33(5)\footnotemark[5]  & 0.30(9)  & 0.27  &  &  0.27(2)  \\        
~~$^{34}$Cl&  0.64(7)\footnotemark[5]  & 0.57(9)  & 0.33  &  &  0.64(4)  \\     
~~$^{38m}$K & 0.64(7)\footnotemark[6]  & 0.59(9)  & 0.33  &  &  0.62(5)  \\           
~~$^{42}$Sc&  0.40(6)\footnotemark[6]  & 0.42(9)  & 0.44  &  &  0.49(4)  \\        
~~$^{46}$V &  0.45(6)\footnotemark[6]  & 0.38(9)  &       &  &  0.43(3)  \\       
~~$^{50}$Mn&  0.47(9)\footnotemark[6]  & 0.35(9)  &       &  &  0.51(4)  \\        
~~$^{54}$Co&  0.61(6)\footnotemark[6]  & 0.44(9)  & 0.49  &  &  0.61(4)  \\
~~$^{62}$Ga&  &  1.26-1.32\footnotemark[7]  &  1.42  &  &  1.38(16)  \\
~~$^{66}$As&  &  1.41-1.63\footnotemark[7]  &  0.78  &  &  1.40(16)  \\
~~$^{70}$Br&  &  1.11-1.41\footnotemark[7]  &  &  &  1.35(21)  \\  
~~$^{74}$Rb&  &  0.91-1.05\footnotemark[7]  &  0.74  &  &  1.43(40)  \\        
\end{tabular}
\end{ruledtabular}
\footnotetext[1]{Both $\delta_{C1}$ and $\delta_{C2}$ are taken from
Towner, Hardy and Harvey\protect\cite{THH77}, except as noted.}  
\footnotetext[2]{Both $\delta_{C1}$ and $\delta_{C2}$ are taken from
Ormand and Brown\protect\cite{OB95}}  
\footnotetext[3]{SGII results from Sagawa, van Giai and
Suzuki\protect\cite{SVS96}}  
\footnotetext[4]{Value of $\delta_C$ from Navr\'{a}til, Barrett, Ormand 
\protect\cite{NBO97}}
\footnotetext[5]{The values of $\delta_{C1}$ are taken from ref.\protect\cite{To89}}
\footnotetext[6]{The values of $\delta_{C1}$ are taken from ref.\protect\cite{Ha94}}
\footnotetext[7]{Ref.\protect\cite{OB95} uses two methods to calculate $\delta_{C2}$
for these cases; to be consistent with other numbers in this column, we quote the
results for Hartree-Fock wave functions.}  
\end{center}
\end{table}

We can now address the question of whether the CKM unitarity problem
might be removed by plausible changes in the calculated structure-dependent
corrections embodied in $\delta_C-\delta_{NS}$.  As can be seen from
Table~\ref{t:previous}, the typical value of $\delta_C-\delta_{NS}$ is of
order 0.5\% for the nine well-known cases currently used in the unitarity test. 
To remove the unitarity problem, the nuclear-structure dependent corrections,
($\delta_C - \delta_{NS}$), would {\it all} have to be raised to around 0.8\%.
Neither the present work nor any previous calculation gives any indication that
such a systematic shift is plausible under any reasonable circumstances.

The structure-dependent corrections have another more impressive credential, one
that is not often appreciated: they are demonstrably effective in bringing the
disparate experimental $ft$-values into agreement with CVC.  If the experimental
$ft$-values were left uncorrected, their scatter would be quite inconsistent with
a single value for the vector coupling constant, $\GV$.  Once corrected, the resulting
$\F t$-values are in excellent agreement with this expectation.  In a very real sense,
it can be said that CVC supports the structure-dependent corrections.  This point will
be amplified in the next section.

\section{The $\F t$-values: Present Status and Future Prospects}
\label{s:future}

With improved calculations for $\delta_C$ and $\delta_{NS}$, we are now in
a position to extract corrected $\F t$-values from the current world data
for the nine well known experimental $ft$-values.  To do so, we follow the same
procedure we have used in the past\cite{Ha90,TH98} to arrive at values for
$\delta_C$ that best represent the results from the two groups that have made
complete calculations: in the present situation that means we use
Table~\ref{t:previous} and take an unweighted average of the results in
column three (Ormand and Brown\cite{OB95}) with those in column six (present
work).  Noting that there is a small systematic difference of 0.08\% between the
two sets of calculations, we remove that difference and then analyze the scatter
of all nine pairs of $\delta_C$ results about their respective averages to obtain
a standard deviation of 0.034\%.  Our adopted $\delta_C$ values appear in the
second column of Table~\ref{finalFt} where they also include the adopted ``statistical"
uncertainty of 0.034\%.  (The ``systematic" uncertainty of $\pm 0.04\%$, obtained from
the average difference between the two calculations of $\delta_C$, need not be
applied until $\GV$ is extracted from the $\F t$-values.)
 
\begin{table}
\begin{center}
\caption{Calculated values for the corrected $\F t$-values based on the adopted
(average) $\delta_C$ values and world-average experimental $ft$ values. 
\label{finalFt}}
\vskip 1mm
\begin{ruledtabular}
\begin{tabular}{llll}
& & & \\[-3mm]
Parent
& \multicolumn{1}{c}{Adopted}
& \multicolumn{1}{c}{Experimental}
& \multicolumn{1}{c}{Corrected}  \\
& & & \\[-3mm]
nucleus\
& \multicolumn{1}{c}{$\delta_C$(\%)\footnotemark[1]}
& \multicolumn{1}{c}{$ft$(s)\footnotemark[2]}
& \multicolumn{1}{c}{$\F t$(s) }  \\[1mm]
\hline
& & &  \\[-3mm]
$T_z = -1$: & & &  \\
~~$^{10}$C &  0.17(3)  & 3038.7(45)  & 3072.7(48)  \\         
~~$^{14}$O &  0.24(3)& 3038.1(18)  & 3069.4(26)    \\
$T_z = 0$: & & &  \\
~~$^{26m}$Al& 0.29(3)  & 3035.8(17)  & 3071.4(22)   \\        
~~$^{34}$Cl&  0.61(3)  & 3048.4(19)  & 3070.6(25)   \\     
~~$^{38m}$K & 0.61(3)  & 3049.5(21)  & 3070.9(27)   \\           
~~$^{42}$Sc&  0.46(3)  & 3045.1(14)  & 3075.7(24)   \\        
~~$^{46}$V &  0.41(3)  & 3044.6(18)  & 3074.4(27)   \\       
~~$^{50}$Mn&  0.43(3)  & 3043.7(16)  & 3072.9(28)   \\        
~~$^{54}$Co&  0.53(3)  & 3045.8(11)  & 3072.1(27)   \\
& & &  \\[-3mm]
\cline{3-4}
& & &  \\[-3mm]
 & & \multicolumn{1}{c}{Average $\F t$} & 3072.2(8)   \\
 & & \multicolumn{1}{c}{$\chi^2/\nu$} & 0.6   \\
\end{tabular}
\end{ruledtabular}
\footnotetext[1]{Average of present results with those of Ormand and
Brown\protect\cite{OB95}; both are listed individually in Table~\ref{t:previous}.
The uncertainties are explained in the text.}  
\footnotetext[2]{Data taken from ref.\cite{TH98}.}  
\end{center}
\end{table}

\begin{figure}[b]
\leavevmode
\epsfxsize=7cm
\hspace{-0.5cm}
\epsfbox{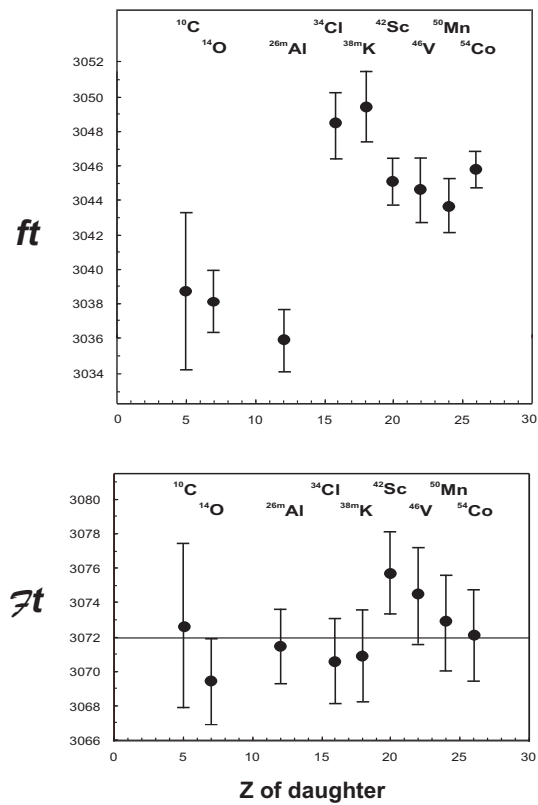}
\caption{Comparison of experimental {\it ft}-values and the corrected $\F t$-values for the nine
well-known superallowed transitions.  This illustrates the effect of the calculated
nucleus-dependent corrections, which change from transition to transition.  (The effect of
$\delta_R^{\prime}$ is virtually the same for all cases.)}  
\label{fig:1}
\end{figure}

The next columns in Table~\ref{finalFt} contain the experimental $ft$-values, which
we have simply taken from ref.\cite{TH98}, and the corrected $\F t$-values, which we
have calculated from Eq.(\ref{Ftnew}) using $\delta_C$ from the first column of this
table, $\delta_{NS}$ from column two of Table~\ref{t:adopt} and $\delta_R^{\prime}$ from
the last column of Table~\ref{radct}.  The average $\F t$-value and the corresponding
$\chi^2$ per degree of freedom also appears at the bottom of the table.
The same information is presented graphically in
Fig.~\ref{fig:1}.  The upper panel shows the uncorrected
experimental $ft$ values and the lower panel the corrected $\F t$ values with the
average indicated by a horizontal line.
Evidently, in these cases, at the current level
of precision the nucleus-dependent corrections act 
very well to remove the considerable ``scatter" that
is apparent in the experimental $ft$-values and is
effectively absent from the corrected
$\F t$-values.  As mentioned already, the consistency of the corrected
$\F t$ values ($\chi^2/\nu = 0.6$) is a powerful validation of the calculated corrections
used in their derivation.

Of course it is only the {\em relative} values of ($\delta_C - \delta_{NS}$)
that are confirmed by the absence of transition-to-transition variations in the
corrected $\F t$-values.  However, $\delta_C$ itself represents
a difference -- the difference between the parent and daughter-state
wave functions caused by charge-dependent mixing.  Thus, the
experimentally determined variations in $\delta_C$ are actually
second differences.  It would be a pathological fault indeed that
could calculate in detail these variations (\ie \ second differences)
in $\delta_C$ while failing to obtain their {\em absolute} values
(\ie \ first differences) to comparable precision.

We have argued that decreasing the radiative correction $\delta_R^{\prime}$
from 1.4\% to 1.1\%, or $\DRV$ from 2.4\% to 2.1\% is unlikely to be
the solution to the CKM unitarity problem; and that there is no
support from calculations for an average increase in the 
nuclear-structure dependent correction, ($\delta_C - \delta_{NS}$),
from 0.5\% to 0.8\%.  We are therefore confident that the unitarity result in
Eq.(\ref{unitarity}), which is unchanged by our new calculations, incorporates
structure-dependent corrections that are correct within their stated
uncertainties.  Nevertheless, these uncertainties are conservatively assigned and,
as we remarked in the introduction, they contribute significantly to the
overall uncertainty of the unitarity test.  There is every reason to continue
to focus on these corrections, both experimentally and theoretically, with a
view to reducing their uncertainties still farther.

One way to do so, of course, would be to increase the precision of the $ft$-values
for the nine cases tabulated in Table~\ref{finalFt} and thus improve the
comparison with CVC that is illustrated in Fig.~\ref{fig:1}.  However, given
the large amount of high-quality data that is already incorporated in these
nine $ft$-values, significant improvements are unlikely in the near term.  A
more promising experimental approach to testing $\delta_C$ is
offered by the possibility of increasing the number of superallowed
emitters accessible to precision studies.  Two series of $0^{+}$
nuclei present themselves: the even-$Z$, $T_z = -1$ nuclei with $18
\leq A \leq 42$, and the odd-$Z$, $T_z = 0$ nuclei with $A \geq 62$. 
The main attraction of these new regions is that the calculated
values of $\delta_C - \delta_{NS}$ for the superallowed transitions are larger,
or show larger variations from nuclide to nuclide, than the values applied
to the nine currently well-known transitions (see Table~\ref{t:adopt}).
In principle, then, they afford a valuable test of the
accuracy of the $\delta_C$ calculations.  It is argued that if the
calculations reproduce the experimentally observed variations where
they are large, then that must surely verify their reliability for the
original nine transitions whose $\delta_C$ values are considerably
smaller.  The calculations reported here, the only complete set available
for all these new cases, should provide a sound basis to which new
experimental data can be compared.

Currently, the greatest
attention is being paid to the $T_z =
0$ emitters with $A \geq 62$, since these
nuclei are being produced at new
radioactive-beam facilities, and their
calculated $\delta_C$ corrections had previously
been predicted to be large
\cite{OB95,SVS96}.  
It is likely, though,
that the required experimental precision will
take some time to achieve.  The decays
of these nuclei are of higher energy and
each therefore involves numerous weak
Gamow-Teller transitions in addition
to the superallowed transition\cite{HT02}.  Branching-ratio 
measurements will thus be very
demanding, particularly with the limited
intensities likely to be available initially for
most of these rather exotic nuclei.  In addition, their
half-lives are considerably shorter than those of the
lighter superallowed emitters; high-precision mass    
measurements ($\pm 2$ keV) for such short-lived activities
will also be very challenging.

More accessible in the short term will be
the $T_z = -1$ superallowed emitters with $18
\leq A \leq 42$.  There is good reason to
explore them.  For example, the calculated
value of ($\delta_C - \delta_{NS}$) for
$^{30}$S
decay,
though smaller than those expected
for the heavier nuclei, is actually 1.13\% -- 
larger than for any other
case currently known -- while $^{22}$Mg
has a low value of 0.51\%.  If such large
differences are confirmed by the measured
$ft$-values, then it will do much to increase
our confidence in the calculated Coulomb
corrections.   To be sure, these decays will
also provide a challenge, particularly in the
measurement of their branching ratios, but the
required precision should be achievable with
isotope-separated beams that are currently
available. 

\section{Conclusions}
\label{conc}

We have presented a new and consistent set of calculations
for the nuclear-structure dependent corrections, ($\delta_C$
- $\delta_{NS}$), required in the analysis of superallowed
$0^{+} \rightarrow 0^{+}$ beta decay.  Twenty transitions have
been included in our calculations, the nine well known ones
already used in the CKM unitarity test, and eleven more that are
likely to be accessible to precise measurements in
the future.  The unitarity test itself is unchanged by our 
calculations, one of several indications we offer that these
corrections are under control within their stated uncertainties.
We have also argued that the structure-independent radiative
corrections are similarly sound.  If the apparent deviation from
unitarity is to be resolved without demanding some extension to
the Standard Model, the only remaining possibility is through undiscovered errors in
$V_{us}$, whose value is currently derived from $K_{e3}$ decay\cite{PDG00,LR84}
and has not been revisited in nearly 20 years.

We have also shown that the uncertainty quoted for the unitarity
test can most effectively be improved by reductions in the uncertainties of
$\DRV$ and ($\delta_C  -  \delta_{NS}$).  We have outlined an experimental
method by which the latter can be improved, and have provided the
full set of calculated corrections that can be tested against experiment.
The stage is now set for a new influx of experimental results on
previously unexplored superallowed transitions, from which the calculated
structure-dependent corrections can be tested and confirmed or
refined.  In either case, the uncertainties should be reduced and the unitarity
test sharpened.  
 
\acknowledgments

The work of JCH was supported by the U. S. Dept. of Energy under Grant
DE-FG03-93ER40773 and by the Robert A. Welch Foundation.
IST would like to thank the Cyclotron Institute of Texas A \& M University
for its hospitality during several two-month summer visits. 

\appendix
\section{Effective Interactions}
\label{s:effi}

In the tables of results presented in the main text, we have only provided
one set of values for each decay studied.  However, for each nucleus,
many calculations were performed with varying choices of effective 
interactions and shell-model spaces.  The error assigned to the adopted values
reflects the spread in the results and our estimate of the uncertainty
in the calculated value based on the quality of the shell-model
calculation.

The choice of an effective interaction is easily made for shell-model
calculations in light nuclei whose principal configurations involve
several valence nucleons away from major shell
closures.  There are well established interactions 
that give excellent fits to spectra.  For $A = 10$, we use
the Cohen-Kurath\cite{CK65} interaction, (8-16)POT, and
for $A = 22$, 26, 30 and 34, we use the universal $s,d$-interaction,
USD, of Wildenthal\cite{Wi84}. For nuclei with $A = 46$, 50 and 54,
we considered two interactions:
the Kuo-Brown $G$-matrix\cite{KB66} as modified by
Poves and Zuker\cite{PZ81} and denoted KB3, and the $fp$-model
independent interaction of Richter \etal \cite{Ri91} and
denoted FPMI3.  For nuclei with $A = 50$ and 54 it was not possible to
perform untruncated calculations in the full $fp$ space;
our calculations only contain $(f_{7/2})^{n-r}
(p_{3/2},f_{5/2},p_{1/2})^r$ configurations with $r \leq 2$.
In this truncated calculation, the spectrum obtained for
$0^{+}$ states in $A = 50$ and 54 is in very poor agreement
with experiment,
a much larger energy gap between the ground state and first
excited $0^{+}$ being obtained.  Thus, we have made further
adjustments to the interaction centroids to obtain a much
improved spectrum in the truncated space. 

For nuclei with $A = 62$, 66 and 70 we considered the model space
$(p_{3/2}, f_{5/2}, p_{1/2})^n$, with $n = A - 56$, which is
based on a closed $f_{7/2}$ shell at the $^{56}$Ni core.  This
model space is the one used by Koops and Glaudemans \cite{KG77}
in their study of nickel and copper isotopes.  We found this
model space, with a modified surface delta interaction (MSDI)
as used in \cite{KG77}, gave acceptable spectra for the
beta-decaying nuclei, with excited $0^+$ states at about
the right excitation energy.

The problem cases were $A=14$, 18, 38, 42 and 74.  In each of
these cases, the experimental excited $0^+$ states are at a much
lower energy than can be obtained in shell-model calculations.
This is symptomatic of the presence of deformed configurations
intruding among the spherical shell-model configurations.
For example, in the $A = 42$
spectrum the lowest-energy states are predominantly
two particles outside a closed $^{40}$Ca core, $\2p$, but
lying low in the spectrum are `intruder' states with a configuration
of four particles and two holes, $\4p2h$.  Mixing between these
configurations must occur, and it is difficult to obtain
the correct degree of mixing with the shell model.  Shell-model
calculations that attempt to mix $\2p$ and $\4p2h$ configurations
encounter what has been called \cite{WB92} the ``$n \hbar \omega$
catastrophe".  The presence of $\4p2h$ configurations depresses
the $\2p$ states, opening up a large energy gap between the
$\2p$ and $\4p2h$ states.  This would be corrected somewhat if
the model calculation included $\6p4h$ states as well, since the
role of the $\6p4h$ states is to depress the $\4p2h$ states.
Thus if the model space is truncated to include only $\2p$ and $\4p2h$
states, the depression driven by the $\6p4h$ states on the $\4p2h$
states is absent.  In an attempt to circumvent this catastrophe
we weakened the cross-shell interactions.  Specifically,
at mass 14, 18, 38 and 42 we used the Millener-Kurath \cite{MK75}
interaction to evaluate the
$\langle 2p | V | 4p \hyphen 2h \rangle$ matrix elements.
We multiplied these matrix elements by a factor, $f$, that ranges
from 0.0 to 1.0.  When $f = 0.0$, there is no mixing between
$\2p$ and $\4p2h$ configurations, and when $f = 0.6$ the ground-state
wave function is approximately $80\% \2p$ and $20\%\4p2h$.
Our strategy was to adjust $f$ so that the excited $0^+$ 
energy is approximately equal to the experimental excitation energy.
We have examined the sensitivity of our results to variations in
$f$ and ensured that the spread of values obtained were within
the assigned errors attributed. 

There are some older interactions that operate in very restrictive
model spaces, but remove the $n \hbar \omega$ catastrophe by allowing
mixing between $\2p$, $\4p2h$ and $\6p4h$ configurations.  These
are the Zuker-Buck-McGrory\cite{ZBM68} interaction, ZBM,
as modified by Zuker \cite{Zu69}, which
uses the $p_{1/2}$, $s_{1/2}$ and $d_{5/2}$ orbitals for
the $A=14$ and 18 nuclei; and the Federman-Pittel\cite{FP69}
interaction, FP, which
uses the $d_{3/2}$ and $f_{7/2}$ orbitals for the $A = 38$ and 42
nuclei.

Finally, at mass 74 there is a related but slightly different problem.
The spectrum in a $(p_{3/2}, f_{5/2}, p_{1/2})^{18}$ model space
gives about the right density of natural-parity states.  The
difficulty is the presence of unnatural-parity states lying low
in the spectrum (for example, $^{73}$Br has a $5/2^+$ at only
280 keV excitation, while $^{75}$Rb has a probable $3/2^+$ at
40 keV).  Further, the excited $0^+$ state in $^{74}$Kr is at only
508 keV, whereas the 
$(p_{3/2}, f_{5/2}, p_{1/2})^{18}$ model calculation puts the
state at 2550 keV.  The influence of the $1g_{9/2}$, $2d_{5/2}$
and possibly $1g_{7/2}$ orbitals is evidently quite strong at 
the end of the $p,f$ shell.  Thus, we have used the following
model space
\be
(p_{3/2}, f_{5/2}, p_{1/2})^{18} + (p_{3/2})^8 (f_{5/2},p_{1/2})^8
(g_{9/2},d_{5/2},g_{7/2})^2 .
\label{pfpgdg}
\ee
Let us call the first term in Eq.~(\ref{pfpgdg}) the
$0 \hbar \omega$
term, and the second term with two nucleons promoted to the $d,g$ shell the
$2 \hbar \omega$ term.  Because of the ``$n \hbar \omega$ catastrophe'',
we again multiply all $\langle
0 \hbar \omega | V |
2 \hbar \omega \rangle$ matrix elements by a factor, $f$, and adjust
$f$ so that the excited $0^+$ state in $^{74}$Kr is reproduced at its
experimental location.  All matrix elements were then calculated with
the MSDI interaction\cite{KG77}.


\begin{thebibliography}{99}

\bibitem{TH98}
I.S. Towner and J.C. Hardy, {\it Proc. of the V Int. WEIN 
Symposium: Physics Beyond the Standard
Model, Santa Fe, NM, June 1998}, edited by P. Herczeg, C.M. Hoffman and 
H.V. Klapdor-Kleingrothaus
(World Scientific, Singapore, 1999) pp. 338-359.

\bibitem{Ha90}
J.C. Hardy, I.S. Towner, V.T. Koslowsky, E. Hagberg and H. Schmeing,
\np {\bf A509}, 429 (1990).

\bibitem{Ab02}
H. Abele, M.A. Hoffmann, S. Baessler, D. Dubbers, F. Gluck, U. Muller,
V. Nesvizhevsky, J. Reich and O. Zimmer, \prl {\bf 88}, 211801 (2002).

\bibitem{PDG00}
D.E. Groom \etal , Eur. Phys. J. C {\bf 15}, 1 (2000).

\bibitem{Si87}
A. Sirlin, \pr D{\bf 35}, 3423 (1987);
A. Sirlin and R. Zucchini, \prl {\bf 57}, 1994 (1986).

\bibitem{JR87}
W. Jaus and G. Rasche, \pr D{\bf 35}, 3420 (1987).

\bibitem{OB85}
W.E. Ormand and B.A. Brown, \np {\bf A440}, 274 (1985).

\bibitem{Ha94}
E. Hagberg, V.T. Koslowsky, J.C. Hardy, I.S. Towner, J.G. Hykawy,
G. Savard and T. Shinozuka,
\prl {\bf 73}, 396 (1994).

\bibitem{THH77}
I.S. Towner, J.C. Hardy and M. Harvey, \np {\bf A284}, 269 (1977).

\bibitem{Si67}
A. Sirlin, \pr {\bf 164}, 1767 (1967).

\bibitem{MS86}
W.J. Marciano and A. Sirlin, \prl {\bf 56}, 22 (1986).

\bibitem{Si94}
A. Sirlin, in {\em Precision Tests of the
Standard
Electroweak Model}, ed. P. Langacker
(World-Scientific,
Singapore, 1994).

\bibitem{JR90}
W. Jaus and G. Rasche, \pr D{\bf 41}, 166 (1990).

\bibitem{BBJR92}
F.C. Barker, B.A. Brown, W. Jaus and G. Rasche,
\np {\bf A540}, 501 (1992).

\bibitem{To92}
I.S. Towner, \np {\bf A540}, 478 (1992).

\bibitem{To94}
I.S. Towner, \pl {\bf B333}, 13 (1994).

\bibitem{To87}
I.S. Towner, Phys. Reports {\bf 155}, 263 (1987).

\bibitem{ASBH87}
A. Arima, K. Shimizu, W. Bentz and H. Hyuga, Adv. in Nucl. Phys.
{\bf 18}, 1 (1987).

\bibitem{BW83}
B.A. Brown and B.H. Wildenthal, \pr C{\bf 28}, 2397 (1986);
B.A. Brown and B.H. Wildenthal, At. Data Nucl. Data Tables
{\bf 33}, 347 (1985);
B.A. Brown and B.H. Wildenthal, \np {\bf A474}, 290 (1987).

\bibitem{NBO97}
P. Navr\'{a}til, B.R. Barrett and W.E. Ormand, \pr C{\bf 56}, 2542 (1997).

\bibitem{Br98}
J. Britz, A. Pape and M.S. Antony, Atomic Data and Nucl. Data Tables
{\bf 69}, 125 (1998).

\bibitem{OB89}
W.E. Ormand and B.A. Brown, \prl {\bf 62}, 866 (1989).

\bibitem{DR85}
W.W. Daehnick and R.D. Rosa, \pr C{\bf 31}, 1499 (1985).

\bibitem{De87}
H. De Vries, C.W. De Jager and C. De Vries,
Atomic Data and Nucl. Data Tables {\bf 36}, 495 (1987).

\bibitem{TH73}
I.S. Towner and J.C. Hardy, \np {\bf A205}, 33 (1973).

\bibitem{To89}
I.S. Towner, in {\it Symmetry Violations in Subatomic Physics},
edited by B. Castel and P.J. O'Donnel
(World Scientific, Singapore, 1989), p. 211.

\bibitem{OB95}
W.E. Ormand and B.A. Brown, \pr C{\bf 52}, 2455 (1995).

\bibitem{SVS96}
H. Sagawa, N. Van Giai and T. Suzuki, \pr  C{\bf 53}, 2163 (1996).

\bibitem{HT02}
J.C. Hardy and I.S. Towner \prl {\bf 88}, 252501 (2002).

\bibitem{LR84}
H. Leutwyler and M. Roos, Z. Phys. {\bf C25}, 91 (1984).

\bibitem{CK65}
S. Cohen and D. Kurath, \np {\bf 73}, 1 (1965).

\bibitem{Wi84}
B.H. Wildenthal, in {\it Progress in Particle and Nuclear Physics},
edited by D.H. Wilkinson (Pergamon Press, Oxford 1984) Vol. {\bf 11},
p. 5.

\bibitem{KB66}
T.T.S. Kuo and G.E. Brown, \np {\bf 85}, 40 (1966).

\bibitem{PZ81}
A. Poves and A.P. Zuker, Phys. Reports {\bf 70}, 235 (1981).

\bibitem{Ri91}
W.A. Richter, M.G. Van Der Merwe, R.E. Julies and B.A. Brown,
\np {\bf A523}, 325 (1991).

\bibitem{KG77}
J.E. Koops and P.W.M. Glaudemans,
Z. Phys. {\bf A280}, 181 (1977).

\bibitem{WB92}
E.K. Warburton and B.A. Brown, \pr C{\bf 46}, 923 (1992).

\bibitem{MK75}
D.J. Millener and D. Kurath, \np {\bf A255}, 315 (1975).

\bibitem{ZBM68}
A.P. Zuker, B. Buck and J.B. McGrory, \prl {\bf 21}, 39 (1968).

\bibitem{Zu69}
A.P. Zuker, \prl {\bf 23}, 983 (1969).

\bibitem{FP69}
P. Federman and S. Pittel, \pr {\bf 186}, 1106 (1969).



\end{thebibliography}
\end{document}